\documentclass[usenatbib,times]{mn2e}
\usepackage{graphicx}
\usepackage{epsf}

\def\msun{{\rm M_{\odot}}}

\def\xte{{\it RXTE}}

\def\H0{{\rm ~km~s^{-1}~Mpc^{-1}}}

\def\msun{M_{\rm \odot}}

\def\deg{^\circ}

\begin{document}

\title[QPO and harmonic in GX~339-4]{Revealing the nature of the QPO and its harmonic in GX~339-4 using frequency-resolved spectroscopy}

\author[Axelsson \& Done]{Magnus Axelsson,$^{1}$\thanks{email: magnusa@astro.su.se}
and Chris Done$^{2}$\\
$^{1}$Department of Physics, Tokyo Metropolitan University, Minami-osawa 1-1, Hachioji, Tokyo 192-0397, Japan\\
$^{2}$Centre for Extragalactic Astronomy, Department of Physics, Durham University, South Road, Durham DH1 3LE, UK\\
}

\date{Accepted --. Received --; in original form --}

\pagerange{\pageref{firstpage}--\pageref{lastpage}} \pubyear{2002}

\maketitle

\begin{abstract}

We use frequency-resolved spectroscopy to examine the energy spectra of the prominent low frequency Quasi-Periodic Oscillation (QPO) and its 
harmonic in GX 339-4. We track the evolution of these spectra as the source makes a transition from a bright low/hard to hard intermediate state.

In the hard/intermediate states, the QPO and time averaged spectra are similar and the harmonic is either undetected or similar to the QPO. By 
contrast, in the softer states, the harmonic is 
dramatically softer than the QPO spectrum and the time averaged spectrum, and the QPO spectrum is dramatically harder than the time averaged 
spectrum. Clearly, the existance of these very different spectral shaped components mean that the time-averaged spectra are complex, as also seen 
by the fact that the softer spectra cannot be well described by a disc, Comptonisation and its reflection. We use the frequency resolved spectra to 
better constrain the model components, and find that the data are consistent with a time-averaged spectrum which has an additional low temperature, 
optically thick Comptonisation component. The harmonic can be described by this additional component alone, while the QPO spectrum is similar to that 
of the hard Comptonisation and its reflection. Neither QPO nor harmonic show signs of the disc component even when it is strong in the time averaged 
spectrum. This adds to the growing evidence for inhomogeneous Comptonisation in black hole binaries. 

While the similarity between the harmonic and QPO spectra in the intermediate state can be produced from the angular dependence of Compton 
scattering in a single region, this cannot explain the dramatic differences seen in the soft state. Instead, we propose that the soft Compton region is 
located predominantly above the disc while the hard Compton is from the hotter inner flow. Our results therefore point to multiple possible mechanisms
for producing harmonic features in the power spectrum. The dominant mechanism in a given observation is likely a function of both inclination 
angle and inner disc radius.

\end{abstract}

\section{Introduction}

Spectral modeling is a powerful tool in understanding the physics of astrophysical sources including accreting black holes. By assigning
observed/modelled spectral components to physical regions of the accretion flow we can learn about its geometry and how it changes as a
response to changes in accretion rate. The most popular physically motivated model for the X-ray spectra of accreting black holes is the
combination of an accretion disc and an inner hot flow where electrons up-scatter the disc photons, giving rise to a comptonised
component. The spectra, especially in the brighter states, additionally require the presence of reflection of the comptonised
photons off some cooler material, presumably the accretion disc. Some combination of these components can generally fit the observed
broadband spectra in all spectral states \citep[for a review, see][]{done07}.

However, the data are increasingly showing that more spectral
components must be present, especially in the brightest low/hard
states. The spectra can be matched by including extreme inner disc
reflection \citep{wil99, fab10}, but alternative
models for an additional component include a warm layer or hot spots
on the disc \citep{dis01}, Compton scattering off an additional
thermal or non-thermal electron distribution \citep{now02,ibr05,yam12}, or a
component from the radio jet \citep{now11}. This illustrates the
degeneracy in models which are possible using only spectral data.

The degeneracies can be (mostly) broken by using additional timing
information. Fast timing studies show that there are energy and
frequency dependent lags across the spectrum, both within the
Comptonisation component and between the disc and Comptonisation
component \citep{miy89, now99, utt11}. The lags within the continuum are best explained if the
Comptonisation is not a single temperature, but is softer/cooler at
larger radii, and harder/hotter at smaller radii. This spatial
dependence of the spectrum of the accretion flow, i.e., the
inhomogeneous nature of the comptonisation region, is crucial to
produce the frequency dependent hard lag (soft lead) seen in the
data. Longer timescale variability is generated at larger radii,
affecting first the softer comptonisation, and then propagates down
the accretion flow to modulate the fastest variability with a harder
spectrum produced in the smallest regions close to the black hole
\citep{miy89,now99,kot01, utt05}.

Since the fast timing results require inhomogeneous comptonisation, this means that the broadband continuum seen in the time averaged
spectrum must be made from multiple Compton components. There may be additional contributions from the jet and/or reflection and/or
inhomogeneous disc emission, but the major Compton continuum cannot be a single component. In previous papers we have used the technique 
of frequency-resolved spectroscopy \citep{rev99} to track the inhomogeneity of the comptonisation region. We used data from the
black hole binary XTE~J1550-564 during its rise to outburst as it made a transition from the bright low/hard to very high state. We found
that the fastest variability was always harder than the time averaged comptonisation component, consistent with this being from the
innermost, hardest part of the flow \citep{axel13}. The strong low-frequency oscillation also showed a very
similar spectrum to that of the most rapid variability, consistent with models where the QPO is from Lense-Thirring (relativistic
vertical) precession of the inner parts of the flow \citep{ing09,adh14}. These results directly show the inhomogeneity of the comptonisation 
component. However, the spectral resolution of the fast timing modes used for these data was not good, with the crucial 8--13~keV data 
binned into a single point.  

Here, we use better spectral resolution fast timing data from a similar transition of the black hole binary GX~339-4, and use the frequency
resolved spectrum of the QPO and its harmonic to constrain the comptonisation components in the time averaged spectrum. As before, the 
QPO spectrum is similar to the time averaged comptonisation. However, we show for the first time that the spectrum of the harmonic can 
be strongly different from that of the QPO. We suggest that there are two major comptonisation regions, a soft one where the corona is 
over the truncated disc, and a hard one in the inner regions with no disc. Both are probably inhomogeneous.  

The broadband spectra then must be composed of a disc, soft Compton (similar to the harmonic) and hard Compton continua (similar to 
the QPO) as well as reflection. Neglecting the soft compton continua as well as the more subtle inhomogeneity in the hard compton continua 
may lead to the more puzzling results from spectral fitting such as the very large reflection fraction \citep{pla14} and extreme 
spin \citep{fab10}.

\section{GX~339-4}

The low-mass X-ray binary GX~339-4 was discovered by OSO-7 in 1972
\citep{mar73}, and already this first report mentioned strong
variability and periods of quiescence. It is a transient source, with
outbursts occurring on average every few years. The compact object is
believed to have a mass above $7\,\msun$, and thus to be a black hole
\citep{mun08}. The system inclination has not been well determined,
but a limit of $\ge40\deg$ is set by assuming the mass of the black
hole is less than $20\,\msun$. Studies of the spectral evolution
suggest the the inclination is not high \citep{mun08}. GX~339-4 is one of
the most studied black hole transients, with extensive coverage by the
{\it Rossi X-ray Timing Explorer} ({\it RXTE}) satellite during the
2002, 2004, 2007 and 2010 outbursts. A radio jet has been observed
from the source \citep{cor00}, categorizing it as a microquasar.

During an outburst, GX~339-4 goes through the canonical states typically exhibited by black hole binaries. It is also the first source 
in which hysteresis was found in the spectral evolution: the transition from hard to soft state during the rising phase occurs at much higher 
luminosity than the soft-to-hard transition during the decline \citep[see, e.g.,][]{now02}. 

The temporal properties have been studied extensively. Following the typical behaviour of black hole binaries, the broad-band
rapid variability is strong in the hard state and suppressed in the softer states. A strong quasi-periodic oscillation (QPO) is seen, especially
during the hard-to-soft transitions, often accompanied by a harmonic. An extensive review of the variability features of GX~339-4 can 
be found in \citet{mot11}.

\section{Data analysis}

In this study we use archival data of GX~339-4 from the Proportional
Counter Array \citep[PCA;][]{jah96} and High-Energy X-ray Timing
Explorer \citep[HEXTE:][]{hexte} instruments onboard the {\xte}
satellite. While there are many {\it RXTE} observations made of
GX~339-4, only a few have both high temporal and spectral resolution,
with modes suitable for timing often having a single energy bin in the
8--13\,keV range. We have therefore gone through the {\it RXTE}
archive in search of the best observing modes.

The selected observations are made in the time period 2007 February
4--14 (MJD 54135 to 54145), and cover four different stages in the
hard to soft evolution. For the PCA data we extracted Standard2
spectra for each observation applying standard selection criteria. A
systematic error of 1 per cent was added to each bin in the
spectra. The energy band used in the spectral modelling was
3--30\,keV. In the case of HEXTE we used Standard Mode spectra from
cluster B (cluster A was not working at this time), considering the
energy range 30--100\,keV. In the softer states the sensitivity does 
not allow us to go to higher energies. In the harder state, we initially 
attempted to extend the data. However, doing so did not 
change our results and we therefore chose to keep the same limit in all 
observations for consistency.

\subsection{Fourier spectroscopy}
\label{fourspec}
To compare the total spectrum to that of the rapid variability, we
also performed frequency-resolved spectroscopy \citep{rev99,rev01} on
the PCA data. Following the approach of \citet{axel13}, we extracted a
light curve for each available channel and constructed a power density
spectrum (PDS). To obtain the contribution from the QPO and harmonic 
in each energy band, we fit them in the PDS using a Lorentzian function 
and integrate this component \citep[see also][]{sob06}. This gives the
relative contribution for each channel, which is finally combined to make an
energy spectrum corresponding to each variability component. As in the case 
of the Standard2 spectra, we added a 1 per cent systematic error to the
frequency resolved spectra.

\section{Spectral modeling}

\subsection{Time averaged spectra}

As described above, the two main components used in explaining the
spectrum of GX~339-4 are the accretion disc and a hot inner flow where
componisation produces high-energy emission. The high-energy photons
can then scatter off the accretion disc, producing a characteristic
Compton reflection component. We model the disc component using {\sc diskbb}, where the main
parameter is the maximum temperature $T_{\rm bb}$. For the hard comptonisation component we 
use the model {\sc nthcomp} \citep{zdz96,zyc99}, parametrized by the asymptotic spectral index
$\Gamma$, the seed photon temperature $T_0$ and the electron
temperature $kT_{\rm e}$. We set $T_0 = T_{\rm bb}$ as can be expected
if the seed photons are the photons from the disc. This assumes that the electrons are thermal, 
though with data extending to higher energies is it clear that there are also non-thermal electrons present
 \citep{joi07,del08,cab09}. However, below 100 keV the data are adequately modeled 
 by thermal Comptonisation, so we use this description here for simplicity. 
To model the Compton reflection we use {\sc xilconv}, a convolution model calculating the 
reflection spectrum using the {\sc relxill} \citep{g14} code for an arbitrary input spectrum.
The main parameters are the relative amplitude of the reflected component, $R$, the inclination 
(which we fix to $50\deg$) and the ionization parameter of the reprocessing matter, $\xi$. Finally 
we include photoelectric absorption, keeping the value of the column density frozen at 
$N_{\rm H}=0.5\times10^{22}$ cm$^{-2}$. 

\begin{figure*}
\includegraphics[width=175mm]{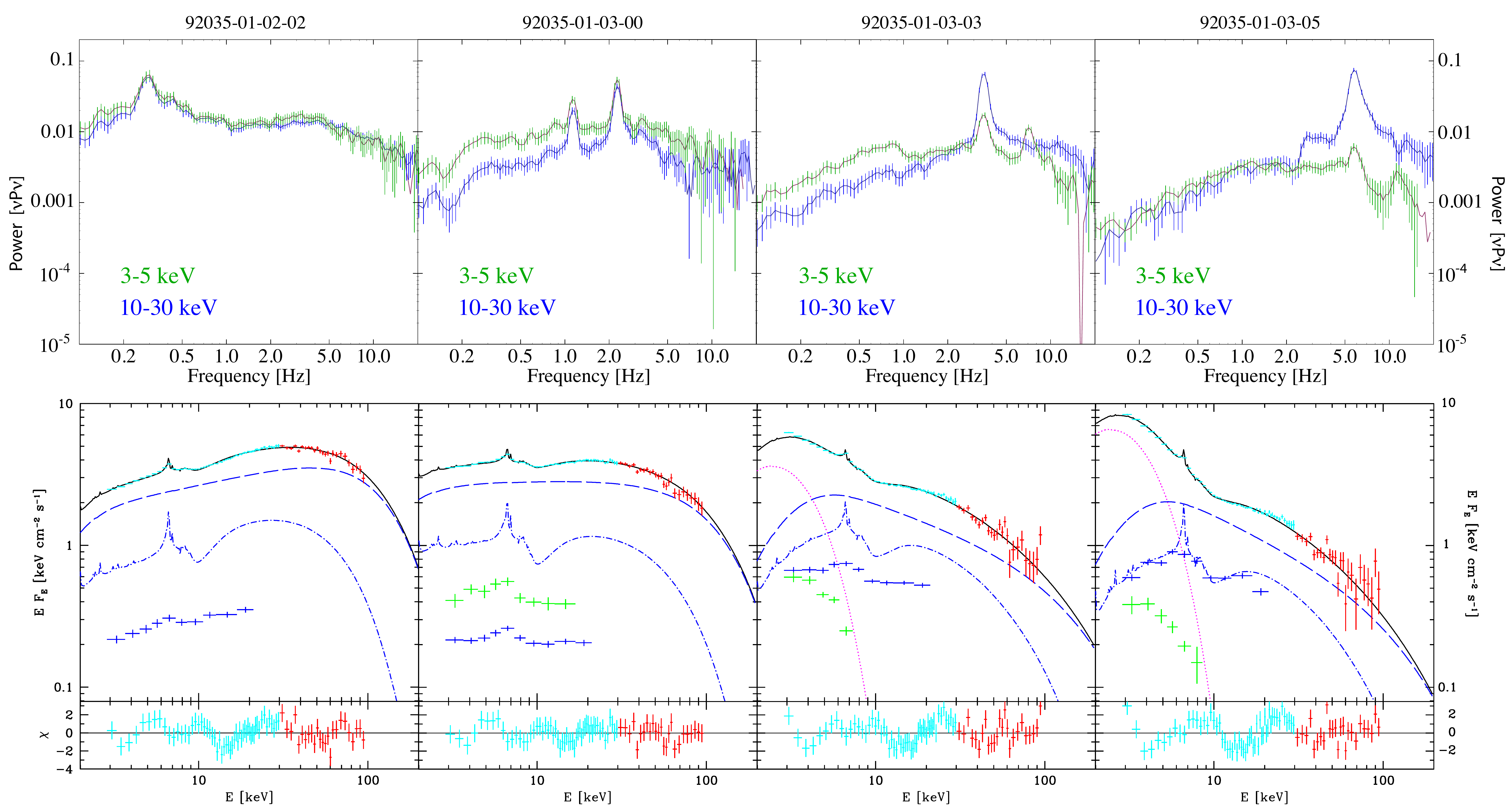}
\caption{Power spectra (upper panels) and fits of our model to the four spectra analysed. The hard-to-soft evolution is clearly seen, as is
the absence of the harmonic in the first observation. The power spectra in the upper panels are shown for two energy bands, 3--5\,keV (green) and
10--30\,keV (blue). In the last two observations, the harmonic is not present in the higher energy band. The lower panels show the PCA (cyan) 
and HEXTE (red) data points overlaid on the total model (solid black line). Also shown are the individual model components: disc blackbody 
(magenta dotted line), main Compton (blue long dashes) and reflection (blue dot-dashed line), as well as the data points of the QPO (blue) and
harmonic (green) spectra. The bottom panels show the residuals.}
\label{contspecs}
\end{figure*}

The results of these fits are shown in Table~\ref{avfits}, and the spectra are plotted in Fig.~\ref{contspecs}, 
together with the power spectrum in two energy bands. While the fits are acceptable for the first two spectra,
in the softer states the model clearly does not provide a good fit. The residuals indicate that the low-energy 
region is most problematic.

Overlaid on the spectra are the data points from the QPO (blue) and harmonic (green) spectra. They 
are quite similar in the intermediate spectrum, but in the softer spectra they are strikingly different. While the spectrum 
of the QPO appears similar to that of the comptonisation component,
the harmonic spectrum is dramatically softer. The presence of the harmonic in the power spectra means
that there has to be a component in the time-averaged spectra matching its spectrum in these observations.
The shape of the harmonic spectrum does not match that of the disc and we therefore test including an additional
soft comptonisation component. It is modelled using {\sc comptt} \citep{tit94}, with the main parameters being the 
input photon temperature (also here tied to $T_{\rm bb}$), the electron temperature $kT_{\rm soft}$ and the optical 
depth $\tau$. While {\sc comptt} further allows switching between disc and sphere geometries, our data do not 
allow us to discriminate between them, and we thus kept the default disc configuration. Though the residuals improve, 
this extra component is not statistically required for the first two spectra (the significance is $\sim3\sigma$); however, 
there is clear improvement in the two softest spectra (significance $>5\sigma$). Although the soft Compton component 
may be present in all spectra, for our modeling we therefore retain it only in the two softest spectra where the shape 
of the harmonic spectrum clearly indicates such a component is needed.

Although the addition of a soft comptonisation component improves the fit, the data of the time-averaged 
spectrum cannot constrain all parameters. This is often the case when using a complex model, and frequently leads 
to degeneracies which are hard to break. In particular the temperatures for the disc and soft Compton components 
are not well determined by the time-averaged spectra alone. As the data only start at 3\,keV the bulk of the disc 
component is not covered and it is therefore not surprising that its temperature is difficult to constrain. In the case 
of the soft comptonisation component, it peaks in a region covered by the data, but where there are many components 
present. Additionally, there is some inherent degeneracy between the electron temperature and optical depth. 
The uncertainties in parameter values are large and some parameters cannot be well constrained. This clearly 
illustrates the difficulty in using multiple spectral components. In spite of this, there 
are still some general trends that can be seen in the data. For instance, the power-law index evolves from hard to
soft as can be expected, and the blackbody temperature is higher in the softer states.

\begin{table*}
\tabcolsep=0.12cm
\begin{tabular}{l l l l l l l l l}
ObsID & $T_{\rm bb}$ & $N_{\rm bb}$  & $\Gamma$ & $kT_{\rm e}$ & $N_{\rm C}$ & $R$ & $\log\xi$ & $\chi^2/dof$ \\
& {\scriptsize (keV)} & & & {\scriptsize (keV)} & & & &\\
\hline
90235-01-02-02 & $0.53_{-0.07}^{+0.04}$ & $0^{+430}$ & $1.79_{-0.02}^{+0.02}$ & $36.9_{-3.4}^{+4.0}$ & $0.46_{-0.07}^{+0.04}$ & $0.61_{-0.11}^{+0.13}$ & $3.11_{-0.07}^{+0.09}$ & 87.0/73 \\  
92035-01-03-00 & $0.60_{-0.08}^{+0.08}$ & $1034_{-328}^{+759}$ & $2.01_{-0.03}^{+0.03}$ & $56_{-11}^{+22}$ & $0.51_{-0.10}^{+0.14}$ & $0.70_{-0.10}^{+0.12}$ & $3.13_{-0.08}^{+0.13}$ & 71.7/73\\
92035-01-03-03 & $0.82_{-0.04}^{+0.05}$ & $1477_{-329}^{+433}$ & $2.41_{-0.02}^{+0.02}$ & $380^{l}$ & $0.30_{-0.02}^{+0.05}$ & $0.86_{-0.15}^{+0.14}$ & $3.08_{-0.07}^{+0.09}$ & 104/73\\
92035-01-03-05 & $0.94_{-0.03}^{+0.03}$ & $1384_{-188}^{+242}$ & $2.48_{-0.04}^{+0.05}$ & $130^{l}$ & $0.18_{-0.02}^{+0.03}$ & $0.72_{-0.16}^{+0.19}$ & $3.13_{-0.09}^{+0.13}$ & 128/73\\
\hline
\end{tabular}
\flushleft
\vspace{-0.2cm}
$^l$ Lower limit.\\
\caption{Parameters derived from fitting the time-averaged spectrum (PCA and HEXTE data) with the model \textsc{wabs}$\times${\sc (diskbb+nthcomp+nthcomp}$\times${\sc xilconv)}. Errors indicate 90\% confidence intervals.}
\label{avfits}
\end{table*}

\subsection{The spectrum of the QPO}

\begin{figure}
\includegraphics[width=80mm]{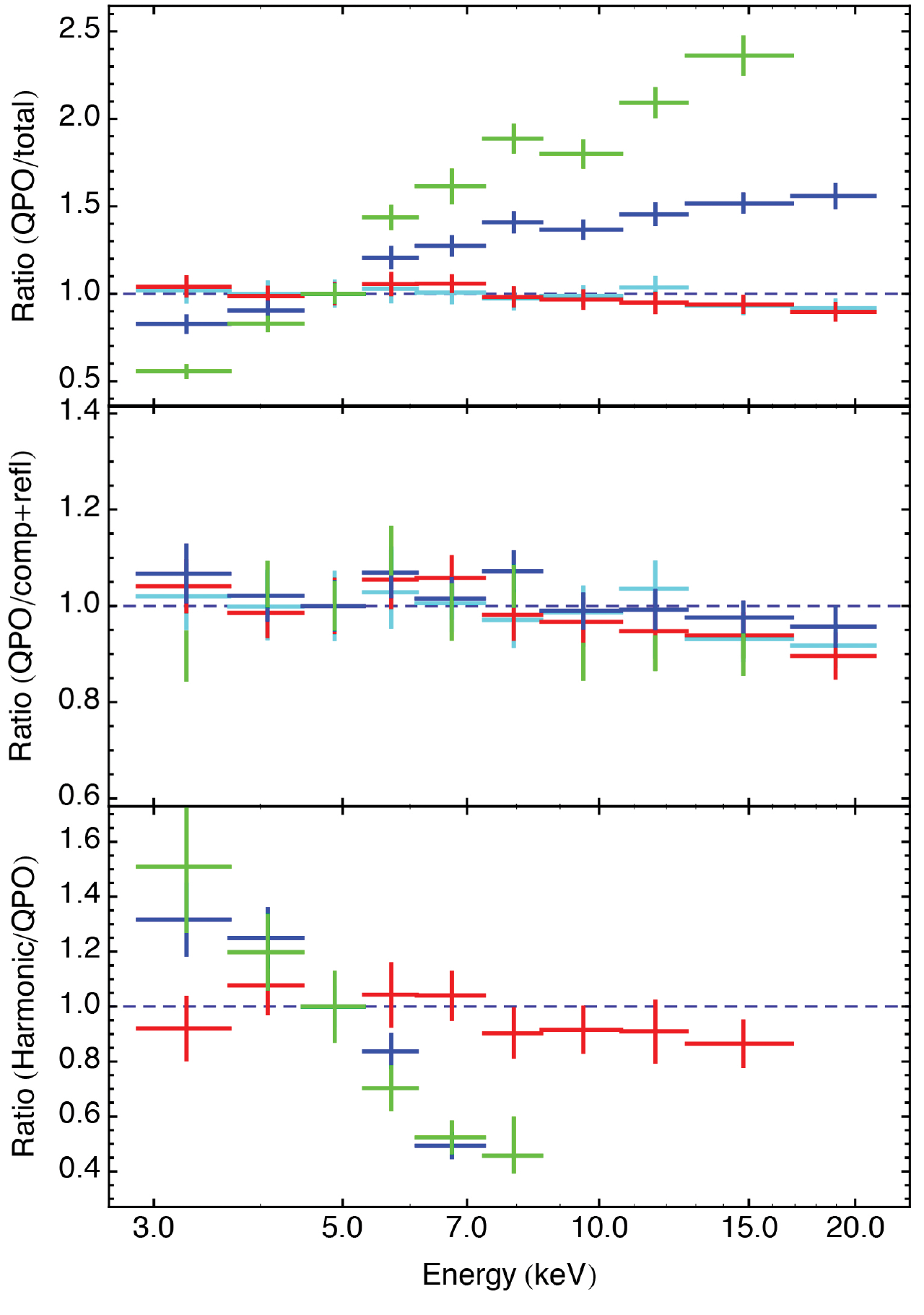}
\caption{Ratio between the QPO and average spectrum (top), QPO and main comptonisation with reflection (middle panel) and 
harmonic to QPO (bottom). All ratios are normalised to 1 at 5 keV. The colours show the evolution during the outburst. From hard 
to soft: cyan, red, blue, green.}
\label{qpocomp}
\end{figure}

The spectrum of the QPO is somewhat different from total spectrum, showing less spectral evolution. 
Studying the ratio between the QPO and time-averaged spectrum (Figure~\ref{qpocomp}, top panel) shows that the QPO 
spectrum is similar to the time-averaged spectrum in the first two observations, but much harder in the last two where the total 
spectrum becomes softer.

We now test how the spectra of the QPO match our model. These spectra have fewer data points and coarser energy 
binning than the average spectra, so it is not possible to constrain the full model using this data alone. In a first step, the 
best-fitting model for the average spectrum is scaled to the level of the QPO spectrum. This provides an acceptable fit in the 
hardest observation, but not for the others. In the softest spectra, it is clear that the strong blackbody and soft Compton component 
prevent the model from fitting the data. 

In order to test for the presence of the disc and soft Compton component in the QPO spectra, we allow the normalisation 
of the blackbody ($N_{\rm bb}$), soft Compton component ($N_{\rm soft}$), spectral index ($\Gamma$) and amount 
of reflection ($R$) to vary in the QPO spectrum. Neither the blackbody nor soft Compton component is statistically required 
in the QPO spectra, even in the softest states when they are very strong in the average spectrum. 

Although the fits converge, we find that even our simplified model is too complex to be uniquely constrained by the data.
The amount of reflection appears to be consistently high in the QPO spectra, with the maximum allowed reflected fraction of 
1.0 within the error range. The spectral index is similar to that of the average spectrum even in the softer states, which seems 
in contradiction with  Fig.~\ref{qpocomp}. A closer look reveals that this is coupled to the strong reflection. The reflection ``hump'' 
at higher energies is balanced by a steep spectral index to give a stronger comptonisation contribution at lower energies. The result 
is a fairly flat overall spectrum, as seen in the lower panels of Fig~\ref{contspecs}. We stress that a weak soft Compton component 
would also give a contribution at low energies; there is indeed a negative correlation between the strength of this component 
and the spectral index.

While we cannot rule out a weak contribution from the disc or soft Compton components, the simplest model which can match 
the QPO spectrum is the combination of main comptonisation and its reflection. We find that these two components alone are able 
to give good fits to all four observations. This is also evident in the middle panel of Fig.~\ref{qpocomp}, showing the ratio of the QPO 
spectrum to the main comptonisation and reflection from the time-averaged spectrum.

\subsection{The spectrum of the harmonic}


The spectra of the harmonic show dramatic differences between the observations. In the low/hard state (ObsID 92035-01-02-02), no
harmonic is present, but the spectra in the other states are shown in the lower panels of Fig.~\ref{contspecs}. The three observations are 
all from different stages of the hard intermediate state, yet the harmonic spectrum varies greatly. In the hardest of the three (ObsID 
92035-01-03-00) the harmonic is stronger than the QPO, yet shows a very similar spectrum. In the two softer observations, the harmonic
is weaker and its spectrum very different from either the average spectrum or the QPO. The greatest change is seen at high energies, 
with the harmonic disappearing in the power spectra above $\sim10$\,keV. At first glance, the harmonic spectrum in the two last observations
appears similar to a blackbody spectrum, although the peak is at higher energy than the disc. We therefore attempt to fit them using a 
single-temperature blackbody, e.g., corresponding to the harmonic arising in a confined region at the inner boundary of the disc. However,
this does not provide a good fit - the spectrum of the harmonic is too broad.

\begin{figure}
\includegraphics[width=85mm]{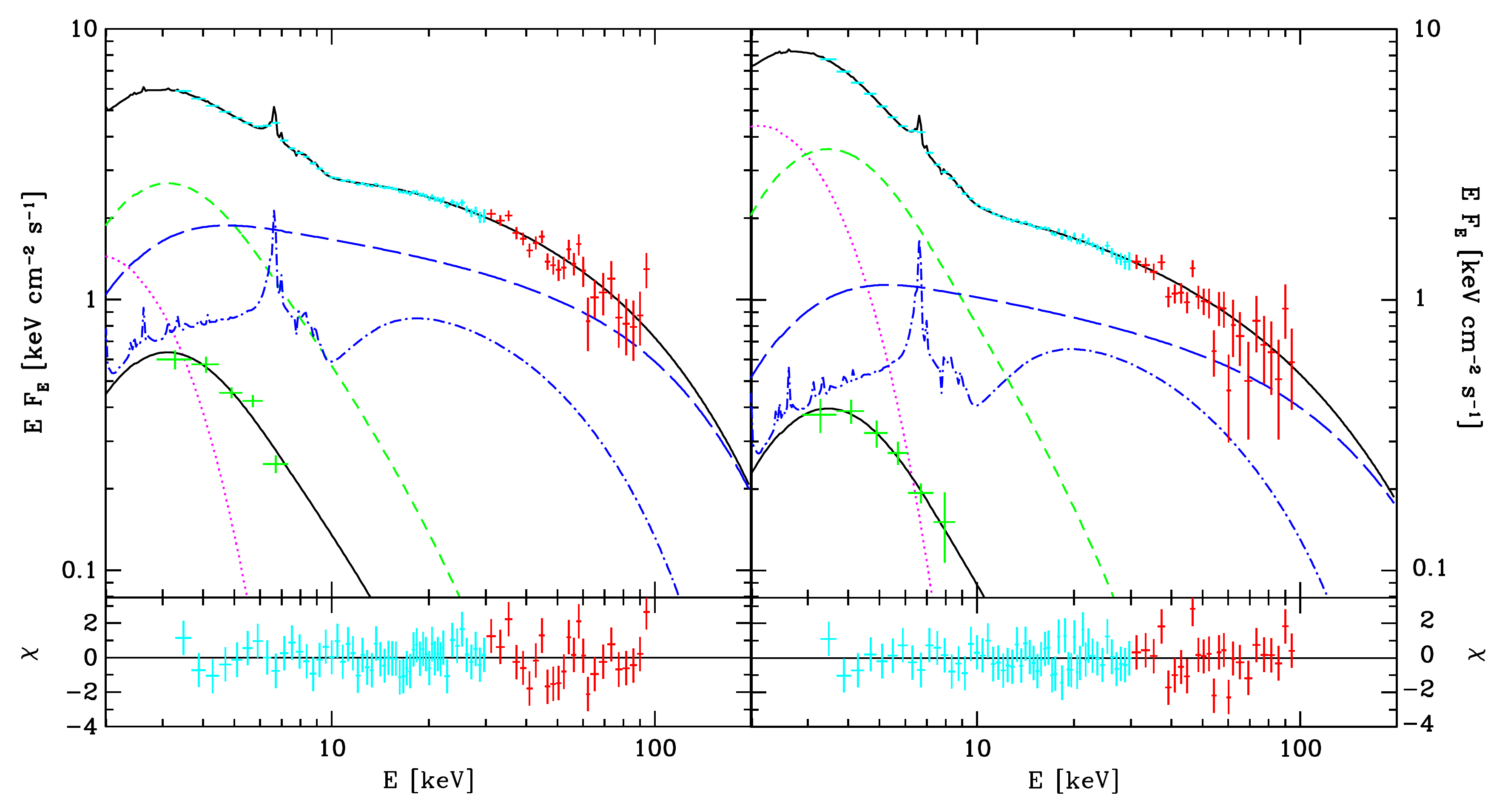}
\caption{Spectral fits for the two softest observations when including a soft comptonisation component in the model. The upper panels show 
the PCA (cyan) and HEXTE (red)  
data points overlaid on the total model (solid black line). Also shown are the individual model components: disc blackbody (magenta dotted line), 
soft Compton (green short dashes), main Compton (blue long dashes) and reflection (blue dot-dashed line). The data points of the harmonic 
(green) spectra are overlaid on the scaled-down soft Compton component (lower solid black line). The bottom panels show the residuals.}
\label{xilfits}
\end{figure}

As the presence of the harmonic spectrum motivated us to include the soft Compton component in the time-averaged spectra of the
softer states, it may provide a means to constrain this component. To test this, the harmonic spectrum is 
fit together with the averaged one (PCA+HEXTE). The model used is the same as before, but the harmonic spectrum is fit using the soft 
comptonisation component only (i.e., {\sc wabs$\times$comptt}). The parameters of this component are tied to be the same in the average 
spectrum, only allowing the normalization to differ. The results are shown in Figure~\ref{xilfits} and the parameters given in 
Table~\ref{harmfits}.

\begin{table*}
\tabcolsep=0.12cm
\begin{tabular}{l l l l l l l l l l l l}
ObsID & $T_{\rm bb}$ & $N_{\rm bb}$  & $kT_{\rm soft}$ & $\tau$ & $N_{\rm soft}$ & $\Gamma$ & $kT_{\rm e}$ & $N_{\rm C}$ & $R$ & $\log\xi$ & $\chi^2/dof$\\
& {\scriptsize (keV)} & & {\scriptsize (keV)}  & & & & {\scriptsize (keV)} & & & & \\
\hline
%

92035-01-03-03 & $0.60_{-0.13}^{+0.08}$ & $2400_{-2400}^{+1210}$ & 7.8$^{f}$ & $1.13_{-0.29}^{+0.26}$ & $0.27_{-0.16}^{+0.27}$ & $2.19_{-0.14}^{+0.11}$ & 39$^{l}$ & $0.44_{-0.12}^{+0.12}$ & $0.75_{-0.15}^{+0.25}$ & $3.05_{-0.10}^{+0.12}$ & 74.0/74 \\

92035-01-03-05 & $0.68_{-0.05}^{+0.05}$ & $4060_{-730}^{+640}$ & 9.5$^{f}$ & $0.78_{-0.18}^{+0.05}$ & $0.26_{-0.07}^{+0.09}$ & $2.19_{-0.08}^{+0.12}$ & 36$^{l}$ & $0.20_{-0.04}^{+0.09}$ & $0.98_{-0.26}^{+0.02}$ & $2.99_{-0.11}^{+0.07}$ & 61.9/75 \\
\hline
\end{tabular}
\flushleft
\vspace{-0.2cm}
$^f$Parameter frozen at this value during error calculations.\\
$^{l}$ Lower limit.
\caption{Parameters derived from fitting the time-averaged spectrum (PCA and HEXTE data) simultaneously with the spectrum of the harmonic (except
for ObsID 92035-01-02-02). The spectrum of the harmonic is fit using only the soft comptonisation. Errors indicate 90\% confidence intervals. Note that 
the reflection fraction was constrained to $R\leq1$ in the fits.}
\label{harmfits}
\end{table*}
 
The fit parameters are in general compatible with those of the average fits alone, but constraining the soft Compton component with the harmonic allows 
more robust fits with smaller uncertainties. However, the degeneracy between electron temperature $kT_{\rm soft}$ and optical depth $\tau$ in the soft 
comptonisation prevent us from constraining both parameters. We therefore freeze the temperature at its best-fitting value when performing the 
error estimations. Having done this, the added data from the harmonic allow us to constrain the remaining parameters, which was not possible using 
time-averaged data alone.

\section{Discussion}

Comparing the power spectra of the two energy bands in Fig.~\ref{contspecs}, in the hard state they are almost identical. The variability here is 
thus more or less independent of energy, and likely we are probing the same component (i.e., the main comptonisation) in both bands. A completely
different picture is instead seen in the softer observations. In these cases, the shapes of the PDS match up until $\sim1-2$\,Hz, and then start to differ. 
This pattern cannot be explained by a single varying component. If the strength of this component is different between the bands, this would
led to a dilution of the variability, and thus a shift in the PDS, but the overall shape would be the same. Neither can the behaviour be explained
by ``pivoting'' of the spectral index, as this would also produce a shift in the relative strength (depending on the location of the pivot point)
without altering the shape of the PDS. Instead, the observed behaviour points to an additional component with a characteristic variability 
timescale of $\sim2$\,Hz. This is too rapid for the cool disc, and indicates the presence of a separate hot/geometrically thick component in 
addition to the main comptonisation. It is thus clear already from these considerations that the comptonisation region is inhomogeneous.

Our result that the QPO spectrum is similar to the main Comptonisation agrees with previous results from XTE~J1550-564, though in GX~339-4 
the QPO is subtly softer than the main Comptonisation whereas in XTE~J1550-564 it is subtly harder. However, the behaviour of the harmonic is very 
different between the two objects. In J1550, the harmonic was always softer than the QPO, but it could be fit with a comptonisation spectrum with the 
same electron temperature as for the QPO and main Comptonisation component \citep{adh14}. By contrast, in GX~339-4, the harmonic seen in the 
softer spectra clearly requires a much lower electron temperature than the main Comptonisation component. 

Precession of the hot inner flow can produce both a QPO and harmonic (from the angle dependence of Compton scattering and the non-spherical 
nature of Compton region) so this predicts that these have spectra which are similar to that of the Comptonisation component \citep{ing09,vel13}. 
Subtle differences in spectra can be produced by the Compton region being inhomogeneous, and the QPO and harmonic being weighted to different radii
\citep{adh14}. However, the dramatic difference of the harmonic in the softest states of GX~339-4 seems too marked to be incorporated into this picture. 
Instead, it suggests that there is a another mechanism to produce the harmonic, and that this mechanism is more evident in GX~339-4 than in J1550. 
Yet the two systems are thought to be very similar in binary parameters \citep{mun08}, with the only difference being inclination angle. J1550 is known 
to be at a rather high inclination angle \citep[$74.7\pm3.7$ degrees;][]{oro11}, while GX~339-4 is viewed at intermediate inclination (as inferred from 
both its low disc temperature: \citet{mun13}, and lack of wind features: \citet{pon12}). Thus we suggest that the visibility of the additional harmonic signal 
depends on inclination, being more evident at low inclination angles. This in turn suggests that it connects to the disc, as is also supported by it being 
seen in the time averaged spectrum mostly in states where there is a strong disc component.

\citet{mot15} find differences between the strength of QPOs as a function of inclination, with different dependance for different types of QPOs. 
However, in our study we see no big differences in the behaviour of the QPO between GX 339-4 and XTE J1550-564 \citep[studied in][]{adh14}. 
The big difference is in the harmonic, which in the harder states of GX 339-4 is similar to J1550, but in the softer states changes drastically. This means 
that inclination alone cannot be the cause for the difference in the harmonic spectrum. Studies comparing the coherence as a function of frequency for 
type C QPOs in several black hole sources \citep{rao10,paw15} have found that there are differences between the different harmonics. Some, like the 
fundamental,  are frequency modulated while others appear amplitude modulated. This could tie into the results found here, indicating more than one 
different mechanism behind the harmonics.

The similarity found between the harmonic spectrum and the soft comptonisation component in the average spectrum of GX~339-4 clearly 
suggests they are related. Yet it is not obvious how such a component could produce the harmonic, or why the other emission components do 
not contribute. If the soft Compton component is the outer edge of the inner flow (the region closest to the disc), it would be expected to overlap
the disc during the softer states. This prevents precession, and thus we would not expect the region to be present in the harmonic (or QPO)
spectra at all. As noted above, the QPO spectra do not require a soft Compton component, suggesting that this region is indeed located further 
out in the flow. 

Another picture could be that the harmonic instead arises \emph{because} of the disc. If the variability is a vertical mode in the inner flow,
the stabilizing disc could in practice cut the wavelength in half and thereby cause the frequency to double. This would occur preferentially
in the outer edge of the hot flow, matching the geometry inferred by, e.g., \citet{yam12}. The QPO picks out the flow which can vertically precess
i.e., the hot inner flow. As this comes past the disc it drives a wave into the softer comptonisation region which bounces off the midplane disc, 
so the harmonic picks out the spectrum of the softer comptonisation region. A doubling of one of the characteristic frequencies in 
Cyg~X-1 was also reported by \citet{axe06} during the state transition, suggesting this could be a common behaviour when the hot inner flow
and accretion disc overlap. Although emission from the soft Compton component should be reflected on the disc, the resulting feature would 
be shifted down in energy and not appear in the {\it RXTE} bandpass, explaining the match between harmonic and direct emission. 

The need for a soft Compton component has also been found in other sources, such as Cyg~X-1. In a systematic study of spectral evolution in 
the hard and intermediate states, \citet{ibr05} found that the optical depth of the soft Compton components decreases for the softer states. 
However, \citet{mak08} find no significant difference in optical depth of their soft Compton component between the hard and hard intermediate 
state. Although the optical depth is lowest in our softest observation, since we only clearly see the component in two spectra we cannot
make any definite statements. 

The geometrical origin of the harmonic, as well as an origin in the inner disc or overlap region, is strongly dependent on viewing angle
and thereby inclination. This could explain the differences seen between GX~339-4 (low to moderate inclination) and J1550 (high inclination). 
In addition, both mechanisms also depend on the inner radius of the accretion disc. Nevertheless, the model proposed here predicts that the 
geometric origin of the harmonic should be strongest in high-inclination sources, whereas the mechanism behind the softer harmonic spectra
(presumably related with the inner disc region) is seen for low and moderate inclination sources. This can be tested with future systematic
studies.

The results also demonstrate that the geometry is likely complex. Comparison of the QPO and average spectra both
here and in \citet{adh14} show that the hard Compton region is inhomogeneous \citep[see also][]{hj15}. It is therefore not difficult to imagine that 
the same would apply also to the soft Compton region. Our simplified models are not sufficient to capture
these details, and the available data are not able to constrain more complex combinations of components. We have assumed a single 
temperature of both the input seed photons as well as the comptonising electrons; in reality, there will be a distribution of temperatures in
both cases. In an overlap geometry the disc temperature observed directly is also likely to be lower than that seen by the comptonising region 
covering the inner regions of the disc. Despite these difficulties, we find that using our model we can build a coherent picture to explain the 
average spectra together with those of the QPO and harmonic, as well as their evolution. This is an important first step towards a physical 
picture of both spectra and variability in black hole binaries.

\section{Summary and Conclusions}

We have studied the spectrum of the QPO and harmonic in GX~339-4 using the {\it RXTE} observations with the best combination
of spectral and temporal resolution, and compared them to the time-averaged spectrum. Our first result, by comparing power spectra in the 
soft and hard bands, is that the comptonising region is inhomogeneous. This is clear evidence for an additional low-energy component 
being present in the spectrum, at least in the softer spectra.

The QPO spectrum is harder than the time-averaged one in the softer states and can be well fit using two components: thermal 
comptonisation together with Compton reflection off an ionized disc. This matches predictions tying its origins to precession of the inner flow.

The drastic change seen in the spectrum of the harmonic indicates that more than one mechanism must be able to produce it. For harder
spectra, the similarity with the QPO spectrum suggests the same geometrical origin as in J1550. For the softer observations the spectrum of 
the harmonic matches the soft Compton component present in the time-averaged spectrum, 
providing a means to constrain this component. The absence of the other spectral components in the data of the harmonic places its origin, 
and thereby that of the additional low-energy component, at the outer edges of the hot flow. We speculate that it is related to the overlap region, 
with the harmonic arising as the disc penetrates the flow. Which of these mechanisms dominate is likely a complex function of inclination
angle and inner disc radius. We propose systematic studies of multiple sources of varying inclination as a means to test this.

Our results clearly show the power of combining temporal and spectral information. By adding the information from the variability,
the different components in the spectrum can be constrained - something which is not possible by spectral fitting alone. A natural extension
is to combine this with phase-resolved techniques (Ingram et al, in prep.), giving information also on possible modulation of components
with orbital phase.

\section*{Acknowledgements}
MA is an International Research Fellow of the Japan Society for the Promotion of Science. CD acknowledges STFC support from grant ST/L00075X/1. 
This work has been partly supported by The Royal Swedish Academy of Sciences through its foundations.


\begin{thebibliography}{}

\bibitem[Axelsson et al.(2006)] {axe06} Axelsson M., Borgonovo L., Larsson S.,\ 2006, A\&A, 452, 975
\bibitem[Axelsson et al.(2013)] {axel13} Axelsson M., Hjalmarsdotter L., Done C.,\ 2013, MNRAS, 431, 1987
\bibitem[Axelsson et al.(2014)] {adh14} Axelsson M., Done C., Hjalmarsdotter L.,\ 2014, MNRAS, 438, 657
\bibitem[Caballero-Garc{\'{\i}}a et al.(2009)]{cab09} Caballero-Garc{\'{\i}}a M.~D., Miller J.~M., Trigo M.~D., et al.\ 2009, ApJ, 692, 1339 
\bibitem[Corbel et al.(2000)]{cor00} Corbel S., Fender R.~P., Tzioumis A.~K., et al.,\ 2000, A\&A, 359, 251
\bibitem[Del Santo et al.(2008)]{del08} Del Santo M., Malzac J., Jourdain E., Belloni T., Ubertini P.\ 2008, MNRAS, 390, 227 
\bibitem[di Salvo et al.(2001)]{dis01} di Salvo T., Done C., {\.Z}ycki P.T., Burderi L., Robba N.R.,\ 2001, ApJ, 547, 1024
\bibitem[Done \& Gierli{\'n}ski(2006)]{dg06} Done, C. \& Gierli{\'n}ski, M., 2006, MNRAS, 367, 659 
\bibitem[Done et al.(2007)]{done07} Done C., Gierli{\'n}ski M., Kubota A., 2007, A\&ARv, 15, 1 
\bibitem[Fabian \& Ross(2010)]{fab10} Fabian A.~C. \& Ross R.~R.\ 2010, SSRv, 157, 167 
\bibitem[Garc{\'{\i}}a et al.(2014)]{g14} Garc{\'{\i}}a J., Dauser T., Lohfink A., et al.\ 2014, ApJ, 782, 76 
\bibitem[Gierli{\'n}ski \& Done(2003)]{gd03} Gierli{\'n}ski M. \& Done C., 2003, MNRAS, 342, 1083 
\bibitem[Hjalmarsdotter et al.(2015)]{hj15} Hjalmarsdotter L., Axelsson, M., Done C., 2015, arXiv: 1502.07135
\bibitem[Ibragimov et al.(2005)]{ibr05} Ibragimov A., Poutanen J., Gilfanov M., Zdziarski A.~A., Shrader C.~R.,\ 2005, MNRAS, 362, 1435 
\bibitem[Ingram et al.(2009)]{ing09} Ingram A., Done C., Fragile P.~C.\ 2009, MNRAS, 397, L101 
\bibitem[Jahoda et al.(1996)]{jah96} Jahoda, K., Swank, J.~H., Giles, A.~B., et al., 1996, Proc. SPIE, 2808, 59
\bibitem[Joinet et al.(2007)]{joi07} Joinet A., Jourdain E., Malzac J., et al.\ 2007, ApJ, 657, 400 
\bibitem[Kohlemainen et al.(2011)]{koh11} Kohlemainen M., Done C., D{\`i}az Trigo M., 2011, MNRAS, 416, 311
\bibitem[Kotov et al.(2001)]{kot01} Kotov O., Churazov E., Gilfanov M.\ 2001, MNRAS, 327, 799 
\bibitem[Makishima et al.(2008)]{mak08} Makishima K., Takahashi H., Yamada S., et al.\ 2008, PASJ, 60, 585 
\bibitem[Markert et al.(1973)]{mar73} Markert T.~H., Canizares C.~R., Clark G.~W., et al.,\ 1973, ApJL, 184, L67  
\bibitem[Miyamoto \& Kitamoto(1989)]{miy89} Miyamoto S. \& Kitamoto S.\ 1989, Nature, 342, 773 
\bibitem[Motta et al.(2011)]{mot11} Motta S., Mu{\~n}oz-Darias T., Casella P., Belloni T., Homan J.\ 2011, MNRAS, 418, 2292 
\bibitem[Motta et al.(2015)]{mot15} Motta S.~E., Casella P., Henze M., Mu{\~n}oz-Darias T., Sanna A., Fender R., Belloni T\ 2015, MNRAS, 447, 2059
\bibitem[Mu{\~n}oz-Darias et al.(2008)]{mun08} Mu{\~n}oz-Darias T., Casares J., Mart{\'{\i}}nez-Pais I.~G.,\ 2008, MNRAS, 385, 2205 
\bibitem[Mu{\~n}oz-Darias et al.(2013)]{mun13} Mu{\~n}oz-Darias T., de Ugarte Postigo A., Russell D.~M., et al.\ 2013, MNRAS, 432, 1133 
\bibitem[Nowak et al.(1999)]{now99} Nowak M.~A., Wilms J., Vaughan B.~A., Dove J.~B., Begelman M.~C.\ 1999, ApJ, 515, 726 
\bibitem[Nowak et al.(2002)]{now02} Nowak M.~A., Wilms J., Dove J.~B.\ 2002, MNRAS, 332, 856
\bibitem[Nowak et al.(2011)]{now11} Nowak M. A., Hanke M., Trowbridge S.N., et al., 2011, ApJ, 728, 13
\bibitem[Orosz et al.(2011)]{oro11} Orosz J.~A., Steiner J.~F., McClintock J.~E., et al.,\ 2011, ApJ, 730, 75
\bibitem[Pawar et al.(2015)]{paw15} Pawar D.~D., Motta S., Shanthi K., Bhattacharya D., Belloni T.\ 2015, MNRAS, 448, 1298 
\bibitem[Plant et al.(2014)]{pla14} Plant D.~S., Fender R.~P., Ponti G., Mu{\~n}oz-Darias T., Coriat M.\ 2014, MNRAS, 442, 1767  
\bibitem[Ponti et al.(2012)]{pon12} Ponti G., Fender R.~P., Begelman M.~C., et al.\ 2012, MNRAS, 422, L11 
\bibitem[Rao et al.(2010)]{rao10} Rao F., Belloni T., Stella L., Zhang S.~N., Li T.\ 2010, ApJ, 714, 1065 
\bibitem[Revnivtsev et al.(1999)]{rev99} Revnivstev M., Gilfanov M., Churazov E., 1999, A\&A, 347, L23
\bibitem[Revnivtsev et al.(2001)]{rev01} Revnivstev M., Gilfanov M., Churazov E., 2001, A\&A, 380, 520
\bibitem[Rothschild et al.(1998)]{hexte} Rothschild R. E., Blanco P. R., Gruber D. E., et al., 1998, ApJ, 496, 538
\bibitem[Sobolewska \& {\.Z}ycki(2006)]{sob06} Sobolewska, M., A. \& {\.Z}ycki, P.T., 2006, MNRAS, 370,405
\bibitem[Titarchuk(1994)]{tit94} Titarchuk L., 1994, ApJ, 434, 313
\bibitem[Uttley et al.(2005)]{utt05} Uttley P., McHardy I.~M., Vaughan S.\ 2005, MNRAS, 359, 345 
\bibitem[Uttley et al.(2011)]{utt11} Uttley P., Wilkinson T., Cassatella P., et al.\ 2011, MNRAS, 414, L60 
\bibitem[Veledina et al.(2013)]{vel13} Veledina A., Poutanen J., Ingram A.\ 2013, ApJ, 778, 165 
\bibitem[Wilms et al.(1999)]{wil99} Wilms J., Nowak M.A., Dove J.B., Fender R.P., di Matteo T.,\ 1999, ApJ, 522, 460
\bibitem[Yamada et al.(2012)]{yam12} Yamada S., Makishima K., Done C., Torii S., Noda H., Sakurai S.,\ 2012, PASJ, 45, 4
\bibitem[Zdziarski et al.(1996)]{zdz96} Zdziarski A. A., Johnson W.N., Magdziarz P., 1996, MNRAS, 283, 193
\bibitem[{\.Z}ycki et al.(1999)]{zyc99} {\.Z}ycki, P., Done, C., Smith D.A., 1999, MNRAS, 309, 561
 



\end{thebibliography}
\end{document}